\documentclass[reprint,amsmath,amssymb,aps,pra]{revtex4-2}

\usepackage{graphicx}
\usepackage{dcolumn}
\usepackage{bm}
\usepackage{amsmath}
\usepackage{amsfonts,amssymb} 
\usepackage{tabularx}
\usepackage{booktabs}
\usepackage{multirow}
\usepackage{color}
\usepackage{siunitx}
\usepackage{tabu}
\usepackage[ruled]{algorithm2e}
\usepackage{subfigure}
\usepackage[T1]{fontenc}
\usepackage{color}
\usepackage{hyperref}

\def\<{\langle}
\def\>{\rangle}
\def\lb{\label}

\def\d{{\mathrm{d}}}

\def\R{{\bf R}}

\def\c{{\bf c}}

\def\s{{\bf s}}

\def\x{{\bf x}}
\def\y{{\bf y}}

\begin{document}

\title{Quantum-Inspired Machine Learning for Molecular Docking}

\author{Runqiu  Shu$^{1,2}$}
\author{Bowen Liu$^{2}$}
\author{Zhaoping Xiong $^3$}
\author{Xiaopeng Cui $^2$}
\author{Yunting Li $^{2,4}$}
\author{Wei Cui$^1$}
\email{aucuiwei@scut.edu.cn}
\author{Man-Hong Yung$^{2,5}$}
\email{yung@sustech.edu.cn}
\author{Nan Qiao$^3$} 
\email{qiaonan3@huawei.com}

\address{$^1$School of Automation Science and Engineering, South China University of Technology, Guangzhou 510641, China}
\address{$^2$Central Research Institute, 2012 Labs, Huawei Technologies}
\address{$^3$Laboratory of AI for Science, Huawei Cloud Computing Technologies Co., Ltd, Guizhou, 550025, China}
\address{$^4$State Key Laboratory of Surface Physics and Department of Physics, Fudan University, Shanghai 200433, China}
\address{$^5$Shenzhen Institute for Quantum Science and Engineering}

\date{\today}

\begin{abstract}
  Molecular docking is an important tool for structure-based drug design, accelerating the efficiency of drug development. Complex and dynamic binding processes between proteins and small molecules require searching and sampling over a wide spatial range. Traditional docking by searching for possible binding sites and conformations is computationally complex and results poorly under blind docking. Quantum-inspired algorithms combining quantum properties and annealing show great advantages in solving combinatorial optimization problems. Inspired by this, we achieve an improved in blind docking by using quantum-inspired combined with gradients learned by deep learning in the encoded molecular space. Numerical simulation shows that our method outperforms traditional docking algorithms and deep learning-based algorithms over 10\%. Compared to the current state-of-the-art deep learning-based docking algorithm DiffDock, the success rate of Top-1 (RMSD<2) achieves an improvement from 33\% to 35\% in our same setup. In particular, a 6\% improvement is realized in the high-precision region(RMSD<1) on molecules data unseen in DiffDock, which demonstrates the well-generalized of our method.
\end{abstract}

\maketitle

\section{Introduction}
Drug development is a complex and costly process that involves extensive research and development efforts. Among various structure-based drug design methods, molecular docking stands out as one of the most commonly utilized techniques for predicting essential parameters such as binding mode and affinity between drug molecules and target proteins~\cite{dockingandscoring, useofmolecular, structurebased, comprehensive, accuracyassessment}. By providing valuable insights into the binding interactions between drug molecules and target proteins, molecular docking plays a crucial role in guiding the design of potent and selective drug candidates, thereby contributing to an improved success rate in drug development~\cite{ageometric, comparisonof}.

Considering molecular docking as an optimization problem, researchers aim to find the optimal solution within a predefined chemical space. However, this task is computationally expensive due to the combinatorial explosion~\cite{simulated, computationalcomplexity}. Traditional molecular docking methods involve extensive sampling of the conformational space of docked molecules, followed by the utilization of a scoring function to evaluate and rank the optimized poses~\cite{autodock, autodockvina, quickvina, quickvina2, smina, glide}. However, the accuracy of docking and the reliability of ranking results are often compromised due to the vast search space and limitations of the scoring functions. Recently, the availability of abundant experimental biological, biochemical, and biophysical data has opened up opportunities to establish a direct relationship between the energetic and structural properties of protein-ligand complexes. This has led to the development of data-driven scoring functions using machine learning techniques~\cite{machine}. In the field of molecular docking, machine learning-based approaches have witnessed rapid advancements, giving rise to a range of sophisticated methods~\cite{gnina, tankbind, equibind, diffdock} that offer notable benefits in terms of speed and blind docking capabilities. However, the interaction energy surfaces between molecules often contain many local minima and flat regions. As a result, deep learning models may become trapped in local optima. This could lead to a decrease in the accuracy of the prediction results.

Recent studies~\cite{moleculardocking, quantummolecular, aquantum} have explored the use of quantum computing or quantum-inspired algorithms~\cite{stateof,quantum-inspired,areview,quantumlloyd,theprospects} in drug design, demonstrating their ability to enhance computational speed and achieve comparable or even greater accuracy compared to classical algorithms. Quantum annealing is an advantageous method for solving combinatorial optimization problems and expediting the drug development process~\cite{recentadvances}. Quantum annealing stands out from conventional methods by employing the principle of quantum adiabatic evolution. It starts from the ground state of an initial Hamiltonian and progressively adjusts the Hamiltonian parameters, enabling a gradual transition of the system towards the target state during the adiabatic evolution. If the minimum energy gap of the Hamiltonian is large enough, the system remains close to the approximation of the initial state throughout the annealing process. As a result, the energy of the Hamiltonian at the end of the annealing process corresponds to the global minimum of the problem Hamiltonian. Quantum annealing relies on specialized quantum hardware, but researchers have also proposed quantum-inspired algorithms that simulate the annealing process using classical hardware. This advancement expands the potential applications of quantum annealing across various contemporary fields~\cite{anapplication,boltzmannsampling,solvingthe}.

In this paper, we propose Quantum-inspired algorithms with a gradient field generated by Diffusion model in Melocular Docking (QDMD) to tackle blind molecular docking tasks. The theoretical framework establishes a connection between the evolution of the quantum-inspired algorithm, specifically the  simulated bifurcation algorithm (SB algorithm)~\cite{bifurcation-based,combinatorial, quantumcomputation,high-performance}, and the score-based generative model. 

The gradient information in the optimization process is derived from the scoring function $s_{\theta}(x)$ generated by the deep learning model. As the SB algorithm operates in discrete space, we also design an encoding algorithm to convert ligand poses and the corresponding gradients into the discrete space.  To compare the optimization results, we also employ a pre-trained confidence model to rank and filter the predictions. In the numerical simulations, we follow the previous experimental setup~\cite{diffdock} and compare it with traditional docking and deep learning based docking methods. The results demonstrate that the combined quantum-inspired algorithm with deep learning surpasses the previous methods under the same settings and improves accuracy from 33.24 \% to 35.26 \% compared to DiffDock. Furthermore, in the recently released ligand data where the model training has not encountered the ligands before, the percentage of RMSD < 1 \AA \;is nearly 6 percentage points higher than that of DiffDock. Overall, the combination of quantum-inspired algorithms and deep learning-based scoring functions shows great promise in molecular docking tasks.

The rest of this paper is organized as follows. In Section \uppercase\expandafter{\romannumeral2} we present the background and related work associated with this paper. In Section \uppercase\expandafter{\romannumeral3}, we present the details of our approach. The feasibility of the method is derived and the specific details of the implementation of the docking task using a quantum-inspired algorithm are described. We compare and analyze the numerical simulation results with other docking methods in Section \uppercase\expandafter{\romannumeral4}. Finally, a summary of the paper is presented.

\section{Background}
\subsection{Molecular Docking}
Molecular docking is a critical process in drug design that involves predicting the binding mode and strength between a ligand and a receptor. In cases where the docking pocket of the protein is unknown, blind docking is employed. Blind docking entails exploring conformations within the entire protein region, leading to a high-dimensional conformational space that encompasses various degrees of freedom such as rotation,  translation and additional conformational flexibility for ligands and proteins. Blind docking poses a significant challenge due to the problem of combinatorial explosion~\cite{dodeep}. 

The molecular docking process typically involves two key steps: conformational space search~\cite{systematicsearch, msdock, applicationsof}, followed by the evaluation of pose quality using a scoring function~\cite{dockingandscoring,soft}. Traditional docking software, such as AutoDock Vina~\cite{autodock, effectivenessanalysis}, utilizes a Monte Carlo sampling technique along with a knowledge-based scoring function and the local optimization. Moreover,enhanced sampling methods and scoring functions have been introduce to QVINA and SMINA based on the traditional paradigm ~\cite{quickvina, smina}.
\begin{figure*}[ht]
    \centering
    \includegraphics[scale=0.5]{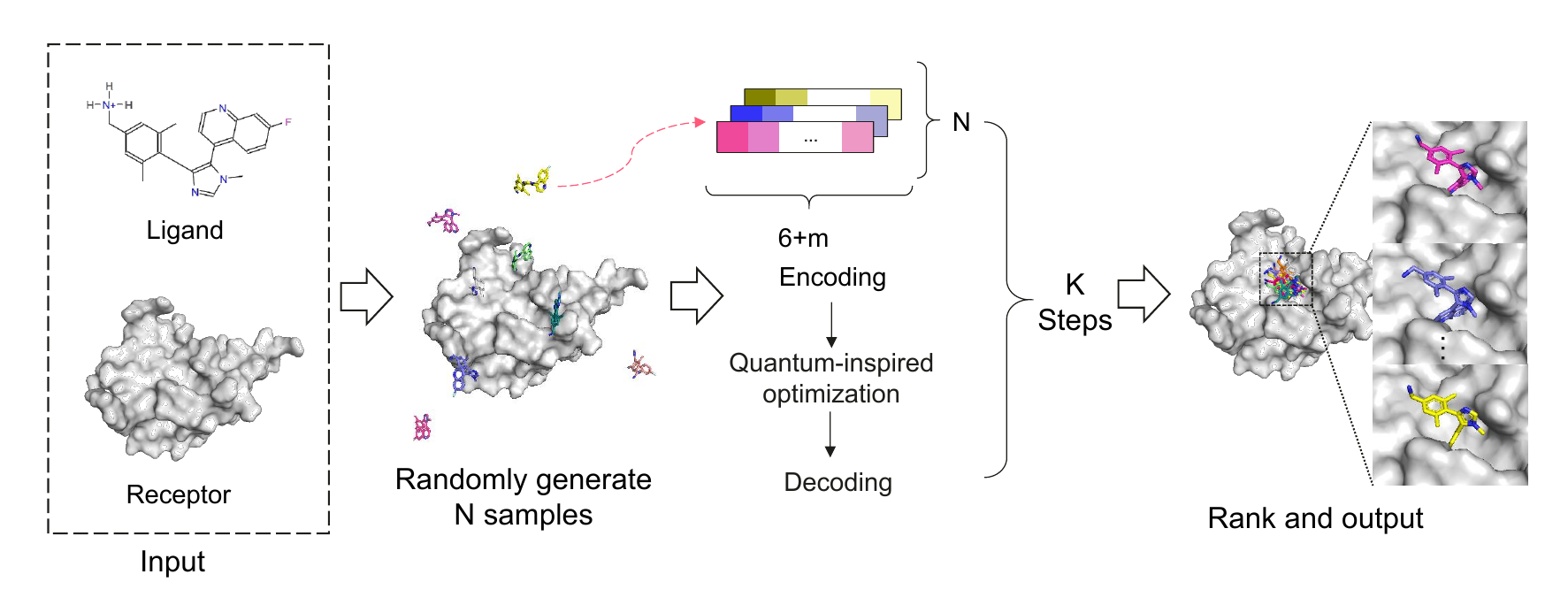}
    \caption{\label{fig1} Overall process of blind molecular docking using QDMD.The main steps of QDMD include random generation of initial positions and conformations, encoding and decoding processes, optimization of Insing values using quantum-inspired algorithm, ranking and filtering of results.}
  \end{figure*}
  
\subsection{Machine learning-based Docking}
In recent years, machine learning-based docking methods have been proposed with the availability of molecular databases and advancements in deep learning models. These methods learn binding patterns between molecules from extensive known data to predict or rank the scoring of poses. GNINA~\cite{gnina} employs a CNN-based deep learning scoring function to rescore generated poses. 
EQUIBIND~\cite{equibind} and TANKBIND~\cite{tankbind} utilize regression-based approaches to predict protein-ligand conformations, significantly accelerating processing speed by several orders of magnitude. Notably, DiffDock~\cite{diffdock} combines diffusion modeling to achieve substantial improvements in blind docking accuracy.

 In DiffDock~\cite{diffdock}, a diffusion generative model~\cite{generative} is introduced to learn the effects of the complex protein-ligand interaction, where the diffusion process takes place in the space of the ligand movement. The model learns a distribution of possible ligand shapes at a crystallographic site, given an initial ligand and a target protein. For the ligand ${\mathbf{\c}}$ in a given protein structure, the probability distribution of $f_\theta(\c)$ is denoted by
\begin{align}\lb{eqn:prob}
	p_\theta(\c)=\frac{e^{-f_\theta(\c)}}{Z_\theta},
\end{align}	
where $\theta$ denotes the parameters and ${Z_{\theta}>0}$ is a normalizing constant dependent on ${\theta}$. However, the constant ${Z_{\theta}>0}$ is difficult to determine. One then can turn to consider ${s_{\theta}(\c)} \equiv \nabla_{\mathbf{\c}} \log p_\theta(\c) =-\nabla_{\mathbf{\c}} f_\theta(\mathbf{\c})$ which is called score function.

According to Ref~\cite{score-based}, Diffdock introduce the diffusion modeling to learn the score function. It involves two processes: diffusion and inverse diffusion. Specifically, 
the diffusion process is based on the Langevin dynamics represented as
$\d \mathbf{c} = g(t)\d\mathbf{w}$, where $g(t)$ is a scaling parameter and $\mathbf{w}$ is the Brownian motion. 
Through training, the deep model acquires the capacity to generalize and estimate the score function of the molecule at various time scales. Once obtained the scored function $s_{\theta}(\mathbf{c})$, the position and conformation of the molecular can be estimated by the inverse diffusion. 
The inverse diffusion represented as
\begin{align}
    \d\mathbf{c} = [ - g^2(t)\nabla_{\mathbf{c}}\log p_{\theta}(\mathbf{c})]\d t +g(t)\d \mathbf{w}.
\end{align}
In addition, DiffDock incorporates a confidence model to rank the final results.

\subsection{Quantum-inspired Algorithm}
Quantum annealing is an optimization algorithm that simulates the annealing process observed in physics to gradually search for the global minimum energy of a system. It is particularly effective in solving combinatorial optimization problems by efficiently search for the global minimum energy state. In the quantum annealing algorithm, 
one typical problem is 
the Ising problem \cite{quantumannealing} where the $N$-spin Ising problem is formulated as 
\begin{align}\lb{eqn:ising}
    H_0(\s)=-\frac{1}{2} \sum_{i=1}^n \sum_{j=1}^n J_{i, j} s_i s_j,
\end{align}
where 
$\s = (s_1, \dots, s_n)\in \{-1,1\}^n$ denotes a spin configuration, and ${J_{ij}}$ is the coupling coefficient between the ${i}$-th and ${j}$-th spins. The configuration probability is given by the Boltzmann distribution
\begin{equation}\lb{eqn:boltzman}
	p_\beta (\s)=\frac{e^{-{\beta}H_0(\s)}}{Z_\beta},
\end{equation}
where $\beta$ is the inverse temperature and $Z_\beta = \sum_{\s\in \{-1,1\}^n} e^{-{\beta}H_0(\s)} $ is the normalization constant similar as the one in Eq.~\eqref{eqn:prob}. 

Inspired by the quantum annealing, the quantum-inspired algorithm is a classical algorithm  solving Ising problem on the classical hardware. In the past decades, several profound quantum-inspired algorithm have been developed to optimize the Ising problem \cite{combinatorial,mathematical}. 
In this paper, we consider the bSB algorithm in Ref.~\cite{combinatorial}. 
It approximate the quantum annealing Hamiltonian by the classical Hamiltonian as follows.
\begin{equation}
    \begin{split}
	H_{\mathrm{bSB}}(\x, \y, t)=\frac{a_0}{2} \sum_{i=1}^n y_i^2+ \frac{a_0-a(t)}{2} \sum_{i=1}^n x_i^2 - \xi_0 H_0(\x),
    \end{split}
\end{equation}
where  $\x = (x_1, \dots, x_n)\in \R^n$ and $ \y = (y_1, \dots, y_n)\in \R^n$ are the position and momentum, $H_0(\x)$ is the Ising energy defined by Eq. \eqref{eqn:ising},
${a(t)}$ is annealing function increased from $0$ to $1$, ${a_0}$ and ${\xi_0}$ are positive constants. The dynamics of $\x$ and $\y$ are given by
\begin{equation}
	\begin{aligned}
		\dot{\x} & =
  \nabla_{\y} H_{\mathrm{bSB}} = a_0 \y, \\
		\dot{\y} & =-
\nabla_{\x}H_{\mathrm{bSB}}=\left(a_0-a(t)\right) \x+\xi \nabla_{\x} H_0(\x),
	\end{aligned}
\end{equation}
where $\nabla_{\x} H_0(\x)$ is the gradient of $H_0(\x)$ with respect to $\x$.
The quantum-inspired algorithm is to solve the Hamiltonian system numerically and take the sign of $\x$ by $\s = {\bf sgn}(\x) \in \{-1,1\}^n$. Then the solution to the Ising problem is given by $\s$. Reader may refer to Ref.~\cite{mathematical} for the mathematical proof on the connection between the Ising problem and the quantum-inspired algorithm. 

\section{Method}
In this section, we will solving the molecular docking by bridging the gap between the diffusion generative model and a quantum-inspired algorithm. 

Inspired by the formal similarity between probability density distributions Eq.~\eqref{eqn:prob} and Boltzmann distributions 
Eq.~\eqref{eqn:boltzman}, we consider quantum-inspired algorithms in conjunction with a scoring-based generative model~\cite{diffdock} and propose the QDMD. As mentioned in the Section \uppercase\expandafter{\romannumeral2}, the optimization of DiffDock is driven by the Langevin dynamics. The gradient field is generated by the score-based function learned from deep learning model. Its inverse process takes the input normal distribution $N(\epsilon, I)$ with sufficiently small $\epsilon$ to a specific distribution $p(\mathbf{c})$. During this process, the scale of the transformation is decreasing, which has similarities with the quantum annealing process. In this sense, we treat the starting normal distribution $N(\epsilon, I)$ as the starting Hamiltonian in the quantum annealing, and $f_{\theta}(\mathbf{c})$ of the final probability distribution $p(\mathbf{c})$ of the docking as the final Hamiltonian. According to the quantum annealing, we evolves the process by time-dependent Hamiltonian equations and realize the docking with the quantum-inspired algorithm on the classical hardware instead of the Langevin dynamics. 

\begin{algorithm}
    \caption{QDMD optimization process}\label{algorithm}
    \KwIn{Protein and Ligand}
    Initialization $\xi_0$ \;
    Define encoding function as $G$;
    
    Randomly generate one ligand pose ${\bf c}_0$ in the conformation space;
    
    \For{k=1 to K}
    {
        Randomly generate $\x_0$ and $\y_0$ in the target space by $\x_0$ and $\y_0$;
        
        Calculate $ \nabla_{\x} F \leftarrow \x_0$; 
        
        \For{i=0 to T}
        {
            $\y_{i+1} \leftarrow \y_i + \left(a_0-a(t)\right) \x_i+\xi_0 \nabla_\x  F$;
            $\x_{i+1} \leftarrow \x_{i} + a_0 \y_{i+1}$;
        }
        ${\bf s}_k=sgn(\x_{T})$\;
        Calculate ${\bf c}_k\leftarrow {\bf c}_{k-1} + G ({\bf s}_k)$\;
    }
    Use the confidence model to rank all results;
    
    \KwOut{Top-5 results ${\bf \bar c}_1, \dots{\bf \bar c}_5$}
    \end{algorithm}

For the sufficiently small $\epsilon$ in the diffusion model, one can take the starting position of each dimension as sufficiently small in the exception sense. It yields that  
we can treat the starting normal distribution $N(\epsilon, I)$ as the starting Hamiltonian, and treat the final probability distribution $f_\theta(\x)$ of the docking as the final Hamiltonian in the quantum annealing. However, quantum-inspired algorithms target discrete variables, where the variables belong to $\{-1,+1\}$. Therefore, we introduce the encoder defined in $(\{-1, 1\}^n)^{m+6}$ and maps to $\R^3 \times \R^3 \times \R^m $ which is the tangent space of $\R^3 \times SO(3) \times SO(2)^m $ by
$G(\s_1, \dots ,\s_{m+6}) =(g(\s_1), \dots, g(\s_{m+6}))$ with 
\begin{equation}
	\begin{split}
		g(\s_l)  =-\sum_{j=1}^n 2^{i-2} \bigg(\prod_{k=1}^{n-j}\left(-s_{lk}\right)\bigg),
	\end{split}
\end{equation}
 where $(\s_1, \dots, \s_{m+6}) \in (\{-1,1\}^{n})^{m+6}$ with $\s_l=(s_{l1}, \dots, s_{ln}) \in \{-1, 1\}^n$. Via the group action $G({\bf s})$, the ligand position and conformation can be updated by the action map $A$ defined in $\R^3 \times SO(3) \times SO(2)^m$ by ${\bf c}_{\bf s} = A(G({\bf s}), {\bf c}_0)$ where ${\bf c}_{\bf s} $ is the new conformation of the ligand.

Abusing the notations, we further expand the discrete domain of the map $G$ from $\{-1,1\}^{n(6+m)}$ to continuous domain $[-1, 1]^{n(6+m)}$ by $G(\x_1, \dots ,\x_{m+6}) =(g(\x_1), \dots, g(\x_{6+m}))$ where
$$g(\x_l)  =-\sum_{i=1}^n 2^{i-2} \bigg(\prod_{k=1}^{n-i}\left(-x_{lk}\right)\bigg),$$ 
where $\x_l = (x_{l1}, \dots, x_{ln}) \in [-1,1]^{n}$.
This encoder bridge the gap between the tangent space of the molecular docking $\R^{6 + m}$ and the quantum-inspired algorithm in $(\{-1,1\})^{n(6+m)}$ and $[-1,1]^{n(6+m)}$. 

Denote the composition of  the action map $A$ with the score function $f_{\theta}$ by $F(\x) = f_{\theta}(A(G(\x) ,{\bf c}_0))$. Now 
we are ready introduce quantum-inspired molecular docking Hamiltonian with the initial conformation ${\bf c}_0$ as the following.
\begin{align}
	H(\x, \y, t) = \frac{a_0}{2} \sum_{i = 1}^N y_i^2+ \sum_{i=1}^N\frac{a_0 - a(t)}{2}x_i^2 + \xi_0 F(\x, {\bf c}_0),
\end{align}
where $\x = (x_1, \dots , x_{N}) \in \R^N$ and $\y = (y_1, \dots, y_N)$, $N= n(6+m)$, and $f_{\theta}(x)$ is defined by Eq.~\eqref{eqn:prob}. 
Therefore, the corresponding Hamiltonian system is given by 
\begin{equation}\lb{eqn:H.x}
	\begin{aligned}
	\dot{\x} & =\nabla_{\y} H=a_0 \y, \\
	\dot{\y} & =-\nabla_{\x} H=\left[a_0-a(t)\right] \x+ \xi_0 \nabla_{\x} F.
    \end{aligned}
\end{equation}
We use the symplectic Euler method to solve Eq.~\eqref{eqn:H.x} numerically.  We then take the sign of $\x$ by $\s = {\bf sgn}(\x) \in \{-1,1\}^{n(6+m)}$ as the binary result. Via the map $G$, the binary result from $(\{-1,1\}^{n})^{m+6}$ can be converted to the continuous tangent space in $\R^{6 + m}$.  Via the group action $A$, we then obtain the final conformation ${\bf c} = A(G(\s), {\bf c}_0)$.

The process of blind molecular docking using QDMD is illustrated in Fig~\ref{fig1}. Initially, the ligand conformation is removed, and a set of $N$ conformations for the ligand is generated using RDKit. The positions and angles of the generated ligands are randomly initialized. To optimize in discrete space, encoding operations, as described previously, are performed. The discrete variables are iteratively optimized using a quantum-inspired algorithm, and the optimized results are recovered through decoding. These steps are repeated for a total of $K$ times. The final output of QDMD is a ranked result.

\section{Numerical Simulation}
\subsection{Benchmarking}
We use the publicly available PDBbind dataset\cite{pdbbind}, which has data on the structure of protein-ligand complexes and their experimentally measured binding affinities. Based on previous work~\cite{equibind}, the dataset was partitioned by time and 363 pairs of protein-ligand complexes from this dataset after 2019 were randomly selected as a test set. In addition, to verify the generalizability of our algorithm, we selected medium-sized ligands with new ligand structures released after October 2022 from the PDBBind dataset for testing. Preprocessing of the test data includes generating protein language model embeddings and generating initial conformations of the ligands using RDKit.

We test our approach against traditional docking methods and machine learning-based docking methods, respectively. Among them, traditional docking methods include QVina-W, SMINA, and 
GLIDE. Deep learning-based docking methods include Equibind and DiffDock. Since all methods except EquiBind generate and rank multiple structures, we report the highest-ranked prediction as the top-1 prediction. And Top-5
refers to selecting the most accurate pose out of the 5 highest ranked predictions. In the experiment, GNINA, SMINA, and EQUIBIND have the same setup as in Ref~\cite{diffdock}. QVINA-W and GLIDE adopt the experimental results of Ref~\cite{diffdock}

To evaluate the generated ligands, we compute ligand RMSD and centroid RMSD separately. 
The ligand RMSD is heavy-atom RMSD (permutation symmetry corrected) between the predicted and the ground-truth ligand when the protein structures are aligned. Centroid RMSD is defined as the distance between the mean 3D coordinates of the predicted ligand atoms and the real bound ligand atoms, indicative of the ability of the model to identify the correct binding region. Following previous work, the ligand RMSD below 2\si{\angstrom}
  is used as a criterion for successful docking.
    
\subsection{Simulation Results}
\begin{table*}[htbp]
	\begin{center}
		\caption{\label{table1}Top-1 statistical results on PDBBind}
		\setlength{\tabcolsep}{3.5mm}
		\begin{tabular}{c|ccccc|ccccc}
			\toprule
			&\multicolumn{5}{c}{\textbf{Ligand RMSD}}&\multicolumn{5}{c}{\textbf{Centroid RMSD}}\\
			&\multicolumn{3}{c}{Percentiles${\downarrow}$}&\multicolumn{2}{c}{threshold${\uparrow}$}&\multicolumn{3}{c}{Percentiles${\downarrow}$}&\multicolumn{2}{c}{thresh${\uparrow}$}\\
			Methods&25th&50th&75th&1\si{\angstrom}&2\si{\angstrom}&25th&50th&75th&1\si{\angstrom}&2\si{\angstrom}\\
			\hline
			QVINA-W&2.5&7.7&23.7&-&20.9&0.9&3.7&22.9&-&41.0\\
			GLIDE (c.)&2.6&9.3&28.1&-&21.8&0.8&5.6&26.9&-&36.1\\
			GNINA&2.4&9.66&25.65&10.47&21.76&0.86&4.91&24.48&26.45&37.74\\
			SMINA&3.73&9.12&23.97&10.45&18.64&1.25&5.05&23.17&22.03&30.79\\
			EQUIBIND&4.33&7.18&14.08&0.28&2.75&1.57&3.42&12.57&13.22&34.16\\
			DiffDock (10)&1.61&4.02&8.20&10.65&30.41&0.56&1.43&3.90&36.84&59.23\\
			DiffDock (40)&1.51&3.90&8.12&13.96&33.24&\textbf{0.52}&1.33&3.89&40.95&62.00\\
			\hline
			QDMD (10)&1.53&3.72&8.05&11.20&32.43&0.56&1.32&3.76&41.44&61.25\\
			QDMD (40)&\textbf{1.42}&\textbf{3.67}&\textbf{7.63}&\textbf{15.05}&\textbf{35.26}&0.54&\textbf{1.28}&\textbf{3.52}&\textbf{42.24}&\textbf{62.83}\\
			\bottomrule
		\end{tabular}
	\end{center}
\end{table*}

\begin{table*}[htbp]
	\begin{center}
		\caption{\label{table2}Top-5 statistical results on PDBBind}
		\setlength{\tabcolsep}{3.5mm}
		\begin{tabular}{c|ccccc|ccccc}
			\toprule
			&\multicolumn{5}{c}{\textbf{Ligand RMSD}}&\multicolumn{5}{c}{\textbf{Centroid RMSD}}\\
			&\multicolumn{3}{c}{Percentiles${\downarrow}$}&\multicolumn{2}{c}{threshold${\uparrow}$}&\multicolumn{3}{c}{Percentiles${\downarrow}$}&\multicolumn{2}{c}{thresh${\uparrow}$}\\
			Methods&25th&50th&75th&1\si{\angstrom}&2\si{\angstrom}&25th&50th&75th&1\si{\angstrom}&2\si{\angstrom}\\
			\hline
			GNINA&1.81&4.76&13.14&12.67&27.82&0.68&1.88&10.55&34.44&51.52\\
			SMINA&1.64&6.74&15.46&16.67&27.97&0.64&3.48&14.13&29.66&39.27\\
			DiffDock (10)&1.41&2.96&5.61&15.06&36.09&0.49&1.16&2.51&44.60&67.40\\  
			DiffDock (40)&1.27&2.83&5.57&18.46&39.67&0.47&1.02&2.32&50.81&70.98\\
			\hline
			QDMD (10)&1.23&2.86&5.59&17.45&38.29&0.48&1.04&2.45&50.83&70.34\\
			QDMD (40)&\textbf{1.15}&\textbf{2.74}&\textbf{5.37}&\textbf{20.48}&\textbf{41.69}&\textbf{0.42}&\textbf{0.95}&\textbf{2.31}&\textbf{51.38}&\textbf{71.44}\\
			
			\bottomrule
		\end{tabular}
	\end{center}
\end{table*}

\begin{figure}[htbp]
    \centering
    \begin{minipage}[t]{1\linewidth}
        \centering
        \label{fig2a}\includegraphics[scale=0.5]{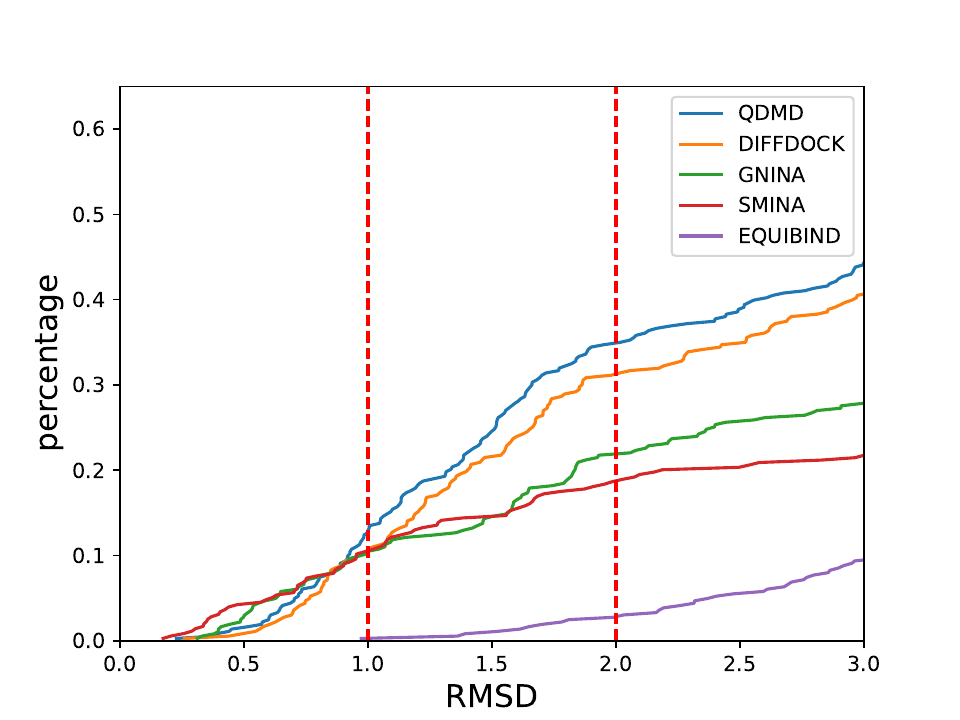}
    \end{minipage}
    \begin{minipage}[t]{1\linewidth}
        \centering
        \label{fig2b}\includegraphics[scale=0.5]{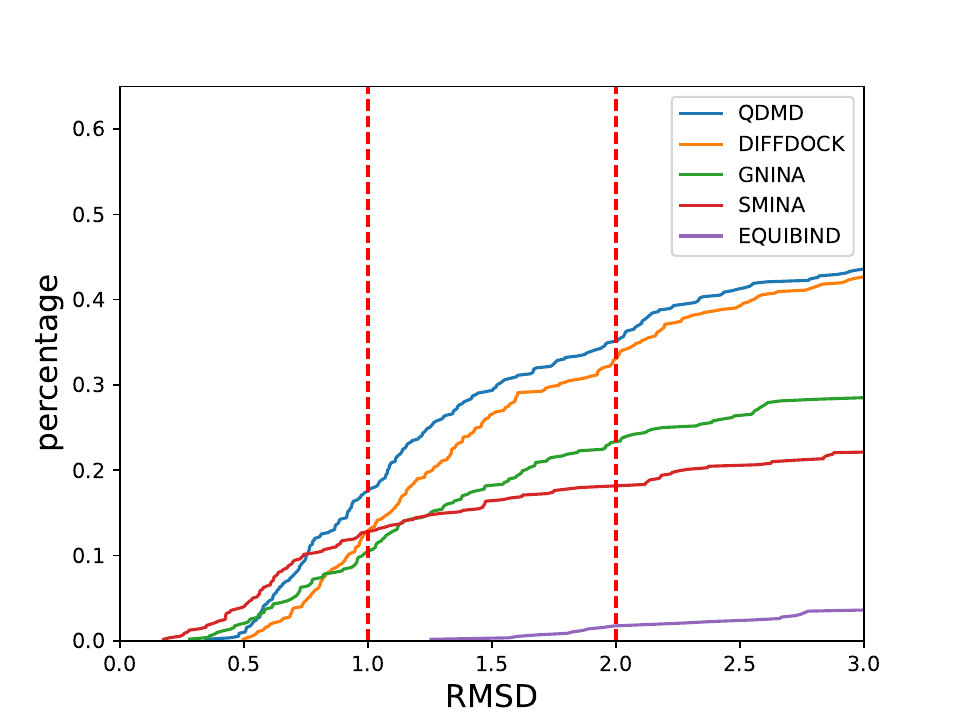}
    \end{minipage}
    \caption{Cumulative RMSD histogram results for different docking methods. The top graph represents the experimental results on 363 pairs of protein-ligand complexes selected based on previous work. The bottom graph represents the experimental results on the newly released ligand dataset.}
    \label{fig2}
    \end{figure}

\begin{table*}[htbp]
	\begin{center}
		\caption{\label{table3}Top-1 statistical results on new curated PDB database}
		\setlength{\tabcolsep}{3.5mm}
		\begin{tabular}{c|ccccc|ccccc}
			\toprule
			&\multicolumn{5}{c}{\textbf{Ligand RMSD}}&\multicolumn{5}{c}{\textbf{Centroid RMSD}}\\
			&\multicolumn{3}{c}{Percentiles${\downarrow}$}&\multicolumn{2}{c}{threshold${\uparrow}$}&\multicolumn{3}{c}{Percentiles${\downarrow}$}&\multicolumn{2}{c}{thresh${\uparrow}$}\\
			Methods&25th&50th&75th&1\si{\angstrom}&2\si{\angstrom}&25th&50th&75th&1\si{\angstrom}&2\si{\angstrom}\\
			\hline
			GNINA&2.29&14.54&26.01&10.35&23.24&1.71&6.40&15.83&11.91&27.54\\
			SMINA&4.64&9.22&19.11&12.63&18.15&1.10&5.73&18.01&17.79&28.29\\
			EQUIBIND&6.36&9.19&13.25&0.00&1.74&3.47&6.16&11.86&2.43&11.48\\  
			DiffDock (10)&1.53&\textbf{4.37}&9.13&11.90&32.73&0.60&1.52&5.11&40.34&57.28\\ 
			\hline
			QDMD (10)&\textbf{1.29}&4.46&\textbf{9.04}&\textbf{17.45}&\textbf{35.00}&\textbf{0.52}&\textbf{1.51}&\textbf{5.10}&\textbf{41.51}&\textbf{57.39}\\
			\bottomrule
		\end{tabular}
	\end{center}
\end{table*}

\begin{table*}[htbp]
	\begin{center}
		\caption{\label{table4}Top-5 statistical results on new curated PDB database}
		\setlength{\tabcolsep}{3.5mm}
		\begin{tabular}{c|ccccc|ccccc}
			\toprule
			&\multicolumn{5}{c}{\textbf{Ligand RMSD}}&\multicolumn{5}{c}{\textbf{Centroid RMSD}}\\
			&\multicolumn{3}{c}{Percentiles${\downarrow}$}&\multicolumn{2}{c}{threshold${\uparrow}$}&\multicolumn{3}{c}{Percentiles${\downarrow}$}&\multicolumn{2}{c}{thresh${\uparrow}$}\\
			Methods&25th&50th&75th&1\si{\angstrom}&2\si{\angstrom}&25th&50th&75th&1\si{\angstrom}&2\si{\angstrom}\\
			\hline
			GNINA&1.71&6.40&15.83&11.91&27.54&0.63&3.16&14.72&33.01&43.36\\
			SMINA&1.62&7.19&11.75&17.79&28.29&0.61&4.24&9.73&30.78&37.19\\
			DiffDock (10)&1.29&\textbf{2.91}&\textbf{6.89}&16.31&38.59&0.55&1.26&\textbf{3.68}&43.99&\textbf{62.45}\\
			\hline
			QDMD (10)&\textbf{1.17}&3.00&7.60&\textbf{21.62}&\textbf{40.06}&\textbf{0.45}&\textbf{1.22}&3.95&\textbf{45.45}&62.32\\
			\bottomrule
		\end{tabular}
	\end{center}
\end{table*}   

According to the described setup, we perform blind docking tests on 363 protein-molecule complexes from PDBBind and 575 complex molecules from a new curated PDB database. A trained confidence model is utilized to rank the optimization results for comparison with DiffDock and other methods. The top-1 and top-5 results on the 363 protein-molecule complexes, averaged over multiple experiments, are reported in Table\ref{table1} and Table\ref{table2}, respectively. 

The results clearly demonstrate that QDMD achieves the highest docking success rate compared to other algorithms in both the top-1 and top-5 cases. Currently, QDMD achieves a docking success rate of 32.43\% with 10 primed samples and reaches 35.26\% with 40 primed samples. Increasing the number of initial samples further enhances the success rate. When compared compared to the DiffDock algorithm, the quantum-inspired algorithm incorporated in QDMD enhances the quality of the solution. It achieves an improvement of 2 percentage points in the overall docking success rate, while operating under comparable time conditions.

Furthermore, to verify the stability and generalizability of the algorithm, we performed experiments using ligands that have not been encountered in the training of current deep learning models. The experimental results on this new dataset are presented in Table~\ref{table3} and Table~\ref{table4}. These results demonstrate that the top-1 success rate of QDMD not only remains unaffected but actually improves when confronted with new data. Compared to traditional methods, QDMD maintains a significant advantage in terms of the success rate. Additionally, QDMD exhibits an improvement of nearly 6 percentage points compared to DiffDock in cases where the RMSD is less than 1. This showcases QDMD's ability to generalize and discover superior solutions within the same gradient field, particularly in high-precision regions. Fig~\ref{fig2} provides a visual representation of the advantages of QDMD in terms of the docking success rate, illustrating its superiority.

\section{Conclusion}
Blind docking poses a significant challenge in the field of molecular docking due to the vast search space and rugged potential energy surface of proteins. Traditional docking methods typically require extensive search time and substantial computational resources, which can lead to challenges in terms of docking accuracy. Molecular docking based on deep learning models may indeed encounter the challenge of falling into local optima due to the complexity of potential energy surfaces. In our study, we employ a quantum-inspired optimization approach to enhance the optimization process. By employing quantum-inspired algorithm, the chances of finding globally optimal solutions are increased by comprehensively and quickly exploring the solution space. Specifically, we establish a link between the quantum-inspired algorithm and the diffusion model. After discretizing the gradient field and molecular pose, we apply the optimization rules of the quantum-inspired algorithm. The results obtained from numerical comparison experiments consistently demonstrate the superior performance of our method over traditional docking and deep learning-based docking for molecular pose and conformation optimization. In summary, our approach, which combines quantum-inspired techniques with deep learning, pushes the boundaries of computational methods in molecular docking. It holds great potential for solving complex molecular optimization problems, and accelerating drug development.

\section*{Acknowledgement}
B. Liu would like to express his sincere thanks to Dr. Choonmeng Lee, Dr. Yao Wang, Dr. Liwei Liu, Dr. Yuxiang Ren and Dr. Xiaozhe Wang for the inspiring discussions on related topics.

\end{document}